\documentclass[reprint,superscriptaddress, aps, amsmath,amssymb,longbibliography, prapplied]{revtex4-2}
\usepackage[T1]{fontenc}
\usepackage{graphicx}
\usepackage{dcolumn}
\usepackage{bm}
\usepackage{color}
\usepackage{textgreek}
\usepackage[english]{babel}
\usepackage{braket, comment}
\usepackage{siunitx}

\usepackage{lastpage}

\usepackage{fancyhdr}
\DeclareUnicodeCharacter{2212}{-}

\usepackage{pgfplots}
\usepackage{pgfplotstable}
\pgfplotsset{compat=newest}

\begin{document}

\title{Low-field all-optical detection of superconductivity using NV nanodiamonds}


\author{Omkar Dhungel} \thanks{These authors contributed equally to this work}
\affiliation{Helmholtz-Institut Mainz, 55099 Mainz, Germany}

\affiliation{Johannes Gutenberg-Universit{\"a}t Mainz, 55128 Mainz, Germany}

\author{Saravanan Sengottuvel} \thanks{These authors contributed equally to this work}
\affiliation{Jagiellonian University, Doctoral School of Exact and Natural Sciences, {\L}ojasiewicza 11, 30-348 Krak\'{o}w, Poland}
\affiliation{%
Institute of Physics, Jagiellonian University, {\L}ojasiewicza 11, 30-348 Krak\'{o}w, Poland}%

\author{Mariusz Mrózek }
\affiliation{%
Institute of Physics, Jagiellonian University, {\L}ojasiewicza 11, 30-348 Krak\'{o}w, Poland}%

\author{Till Lenz}
\affiliation{Helmholtz-Institut Mainz, 55099 Mainz, Germany}
\affiliation{Johannes Gutenberg-Universit{\"a}t Mainz, 55128 Mainz, Germany}
\affiliation{GSI Helmholtzzentrum f{\"u}r Schwerionenforschung GmbH, 64291 Darmstadt, Germany}

\author{Nir Bar-Gill}
\affiliation{%
Institute of Applied Physics, Hebrew University, Jerusalem 91904, Israel}%
\affiliation{%
Racah Institute of Physics, Hebrew University, Jerusalem 91904, Israel}

\author{Adam M. Wojciechowski}
\affiliation{%
Institute of Physics, Jagiellonian University, {\L}ojasiewicza 11, 30-348 Krak\'{o}w, Poland}%

\author{Arne Wickenbrock}
\affiliation{Helmholtz-Institut Mainz, 55099 Mainz, Germany}
\affiliation{Johannes Gutenberg-Universit{\"a}t Mainz, 55128 Mainz, Germany}
\affiliation{GSI Helmholtzzentrum f{\"u}r Schwerionenforschung GmbH, 64291 Darmstadt, Germany}

\author{Dmitry Budker} \email{budker@uni-mainz.de}
\affiliation{Helmholtz-Institut Mainz, 55099 Mainz, Germany}
\affiliation{Johannes Gutenberg-Universit{\"a}t Mainz, 55128 Mainz, Germany}
\affiliation{GSI Helmholtzzentrum f{\"u}r Schwerionenforschung GmbH, 64291 Darmstadt, Germany}
\affiliation{Department of Physics, University of California, Berkeley, USA}

\date{\today}


\begin{abstract} 
Nitrogen-vacancy centers in nanodiamond offer a microwave-free, noninvasive platform for probing superconductors via near zero-field cross-relaxation magnetometry. We demonstrate this by depositing nanodiamonds on YBa$_2$Cu$_3$O$_{7-\delta}$ thin films to measure critical parameters: transition temperature and penetration field. 
This method leverages nanodiamond fluorescence modulation as a result of magnetic field variation with 1\,mT amplitude to observe the Meissner effect and field scans to measure the penetration field. The approach is minimally invasive and can be applied to superconducting samples with rough surfaces, facilitating the study of flux vortices and critical phenomena in complex geometries.
\end{abstract}
\maketitle 

\section{Introduction}
Recent advances in quantum sensing technologies, particularly those utilizing nitrogen-vacancy (NV) centers in diamond, have opened up new avenues for studying condensed matter systems at the nanoscale \cite{rovny_nanoscale_2024, christensen_2024, bhattacharyya_imaging_2024, yip_measuring_2019}. These solid-state quantum sensors offer unique advantages, including quantitative and non-invasive measurements, nanoscale spatial resolution, and operation over a wide temperature range. In recent years, NV centers have been employed to investigate a range of phenomena in superconductors, such as the Meissner effect, critical parameters, and vortex dynamics \cite{bouchard_detection_2011,joshi_measuring_2019,acosta_color_2019,nusran_spatially-resolved_2018, waxman_diamond_2014}. Diamond particles and bulk diamonds were used to investigate the Meissner effect in high-temperature superconductors, revealing critical parameters such as the lower and upper critical fields\,\cite{ho_studying_2024}. NV centers are utilized to map the microwave magnetic field distribution on ultrathin superconducting films, providing insights into the AC Meissner effect and vortex-antivortex dynamics\,\cite{xu_mapping_2019}. The vortex dynamics and critical currents in superconductors under high pulsed magnetic fields were explored, revealing the influence of the rate of change of the magnetic field on current-voltage curves\,\cite{leroux_dynamics_2019}. 
Quantitative nanoscale vortex imaging of cuprate superconductors was achieved by using both confocal and widefield magnetometry\,\cite{thiel_quantitative_2016, schlussel_wide-field_2018}. These studies collectively highlight the potential of NV-based quantum sensing techniques in probing exotic features and dynamics in superconducting materials with high spatial and temporal resolution. 

\begin{figure*}[t]
    \includegraphics[width=0.95\textwidth]{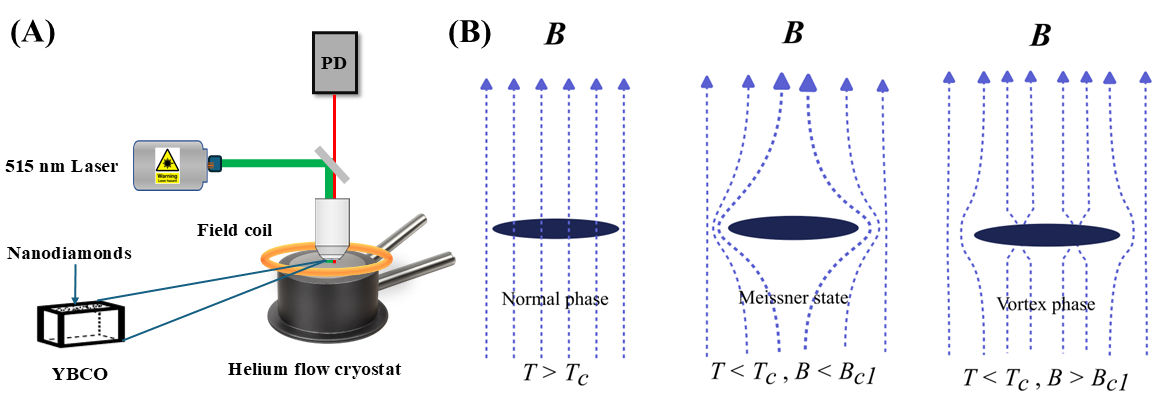}
    \caption{Schematic of the low-temperature widefield magnetic imaging setup and three different phases of high T$_c$ superconductor. (A) NDs of 140\,nm size are deposited on top of YBCO, which is placed on the sample stage of the cryostat. The green laser light is directed onto the NDs and red fluorescence is collected on a photodiode, selected with a flip mirror. A signal generator is used to produce a square-wave signal which is fed into the power supply (not shown) of the magnetic coil as well as to the lock-in amplifier as a reference. (B) Schematic representation of the three characteristic states of a superconductor. In the normal state, magnetic field lines penetrate uniformly through the material. In the Meissner state (below the critical temperature and field), the superconductor exhibits perfect diamagnetism by expelling magnetic flux. In the vortex (mixed) state of a type‑II superconductor, magnetic flux penetrates the material as quantized vortices, each comprising a normal core encircled by superconducting currents.}
    \label{fig:setup}
\end{figure*}

A significant portion of NV-center based superconductor research is based on microwave-induced optically detected magnetic resonance (ODMR) and advanced pulsed methods such as spin-lattice relaxation measurements ($T_1$) and Hahn-echo ($T_2$) measurements\,\cite{monge_spin_2023, nishimura_wide-field_2023}. However, these approaches can present difficulties when investigating a system where high-power microwaves could change the local properties of the sample, or where implementing such control is problematic\,\cite{wickenbrock_magnetic_2014,jensen_cavity-enhanced_2014}, or in scenarios where the use of microwaves is intrusive. These limitations prompt the exploration of alternative sensing modalities that avoid these potential issues. Recently, a microwave-free method was introduced to detect the Meissner state in thin-film superconductors\,\cite{paone_all-optical_2021}. Also, NV magnetometry schemes that operate at near-zero fields without microwave radiation were proposed and demonstrated\,\cite{pellet-mary_relaxation_2023, anishchik_low-field_2015, akhmedzhanov_magnetometry_2019, sengottuvel_microwave-free_2024}.

In this work, we employ a microwave-free all-optical near-zero-field cross-relaxation magnetometry scheme \cite{dhungel_near-zero-field_2024,dhungel_near_2024, sengottuvel_microwave-free_2024} incorporating the so-called "salt-and-pepper" technique \cite{dhungel_near-zero-field_2024}, in which nanodiamonds are sprinkled over the interrogated surface. Cross-relaxation in NV centers arises from resonant dipolar coupling between NV spins and nearby paramagnetic impurities. When the energy splitting of NV sublevels matches that of another spin system, efficient energy exchange enhances the NV spin-lattice relaxation rate. This effect is most substantial near zero magnetic field, where NV orientations are degenerate. By sweeping an external field and monitoring NV relaxation, these resonances appear as dips in fluorescence, forming the basis of cross-relaxation magnetometry. Variations in the position or amplitude of these features directly reveal changes in the local magnetic or spin environment.

Moreover, nanodiamonds offer key advantages for probing condensed matter systems due to their sub-micron dimensions, which allow for minimal standoff distances and proximity to target surfaces, as well as the flexibility to deposit them on textured surfaces and even on fragile materials. This proximity is critical for detecting weak, spatially localized magnetic fields, such as those associated with vortices or edge currents in superconductors \,\cite{thiel_quantitative_2016, schlussel_wide-field_2018}. Together, cross-relaxation magnetometry with NV-NDs makes them a versatile platform for local, non-invasive magnetic measurements in superconducting and other condensed matter systems. This nondestructive and non-perturbative approach offers a microwave-free alternative for probing superconducting materials, potentially providing new insights into their properties and offering potential advancements in superconductivity research.

\section{Experimental technique}
In this work, we investigate the fluorescence modulation of NV centers in NDs placed in proximity to a superconductor when subjected to an external modulated magnetic field. We used commercially available 140\,\si{\nano\meter} carboxylated fluorescent NDs from Adamas Nanotechnologies. We use a YBCO superconductor thin film from Ceraco with a thickness of 200\,nm, which was deposited on an Al$_{2}$O$_{3}$ substrate (5 × 5 × 0.5\,mm$^3$). Two sample types were prepared by dropcasting an aqueous ND suspension onto (i) a transparent cover glass and (ii) the surface of a superconductor, followed by ambient air drying. The samples were then placed in a sample holder mounted on top of the cold finger of a Janis ST-500 helium flow cryostat. To ensure efficient thermal contact, a vacuum-compatible varnish was applied between the sample and the holder.

Fluorescence measurements were carried out using a custom-built widefield microscope as illustrated in Fig.\,\ref{fig:setup}(a). Excitation was provided by a 515 nm Toptica iBeam Smart laser with an output power of $\approx$10\,mW. A dichroic mirror was used to direct the collimated laser beam onto a 50x Mitutoyo long-working-distance microscope objective (NA 0.45). NV fluorescence was collected through the same objective and detected by a photodiode. The region of interest (ROI) was approximately 120\,\si{\micro\meter}$\times$\,120\,\si{\micro\meter} on the nanodiamond layer. A circular coil positioned above the cryostat generated a bias magnetic field perpendicular to the sample plane. The NV magnetometer was calibrated using a test magnetic field and the coil-generated field was cross-verified by ODMR measurements with a {111}-cut bulk-diamond sample containing $~$3\,ppm NV centers. This calibration ensured the accuracy of the magnetic field measurements.

For identifying edge effects related to the magnetic response at the periphery of the superconductor, we examined the cross-relaxation feature during forward and backward sweeps of the external magnetic field over a range of temperatures. The green laser was focused on a spot with a diameter of approximately 120\,$\mu$m, positioned such that approximately 60\,$\mu$m extended into the interior from the sample edge. This arrangement ensured enhanced sensitivity to the magnetic-field distribution in the near-edge region.

For a superconducting YBCO thin film of thickness $d = 200\,\mathrm{nm}$, 
the magnetic stray field of a single perpendicular vortex can be described, 
to good approximation, by the monopole model \cite{Brandt1988,Carneiro2000}. 
In this approach, the vortex is represented as a magnetic monopole carrying 
one flux quantum $\Phi_{0} = h/2e \approx 2.07\times 10^{-15}\,\mathrm{Wb}$, 
located at an effective depth $z_0$ below the surface. The parameter $z_0$ 
is set by the electrodynamic screening length of the superconductor, which lies 
between the London penetration depth $\lambda_{ab}$ and the Pearl length 
$\Lambda = 2\lambda_{ab}^{2}/d$ \cite{London1935,Pearl1964,Tinkham1996}. 
Here, $\lambda_{ab}$ is the London penetration depth for screening currents in 
the superconducting $ab$-plane (for YBCO typically 
$\lambda_{ab} \approx 150$$\,\mathrm{nm}$ at low temperature), and it 
characterizes the exponential decay length of magnetic fields inside the 
superconductor. The Pearl length $\Lambda$ becomes relevant in thin films and 
represents the effective two-dimensional screening length for vortices. For the 
given film thickness and material parameters, one expects 
$z_{0} \approx 200$--$400\,\mathrm{nm}$.

In this model, the out-of-plane magnetic field detected by NDs 
positioned at a height $h$ above the superconductor decays approximately as
\begin{equation}
    B_{z}(r=0,h) \;\approx\; \frac{\Phi_{0}}{2\pi}\,\frac{1}{(h+z_{0})^{2}}\,,
\end{equation}
with the corresponding spatial gradients scaling as $|\nabla B| \sim (h+z_{0})^{-3}$. The sensor--sample distance $h$ is determined by the ND size and geometry: for individual NDs of diameter $\sim 140\,\mathrm{nm}$, the distance between NVs inside 
NDs and the sample is shorter than the size of the NDs; however, when NDs agglomerate into clusters, this distance can increase to several hundred nanometers or even micrometers.  Thus, agglomeration strongly reduces the magnetic signal amplitude and diminishes spatial resolution by averaging the larger footprint of the cluster. Suppressing ND agglomeration and creating a homogeneous distribution of ND across the field of view is therefore crucial for maximizing sensitivity \cite{Kirtley2010,thiel_quantitative_2016}.

\begin{figure}
    \centering
    
    \includegraphics[width=0.47\textwidth]{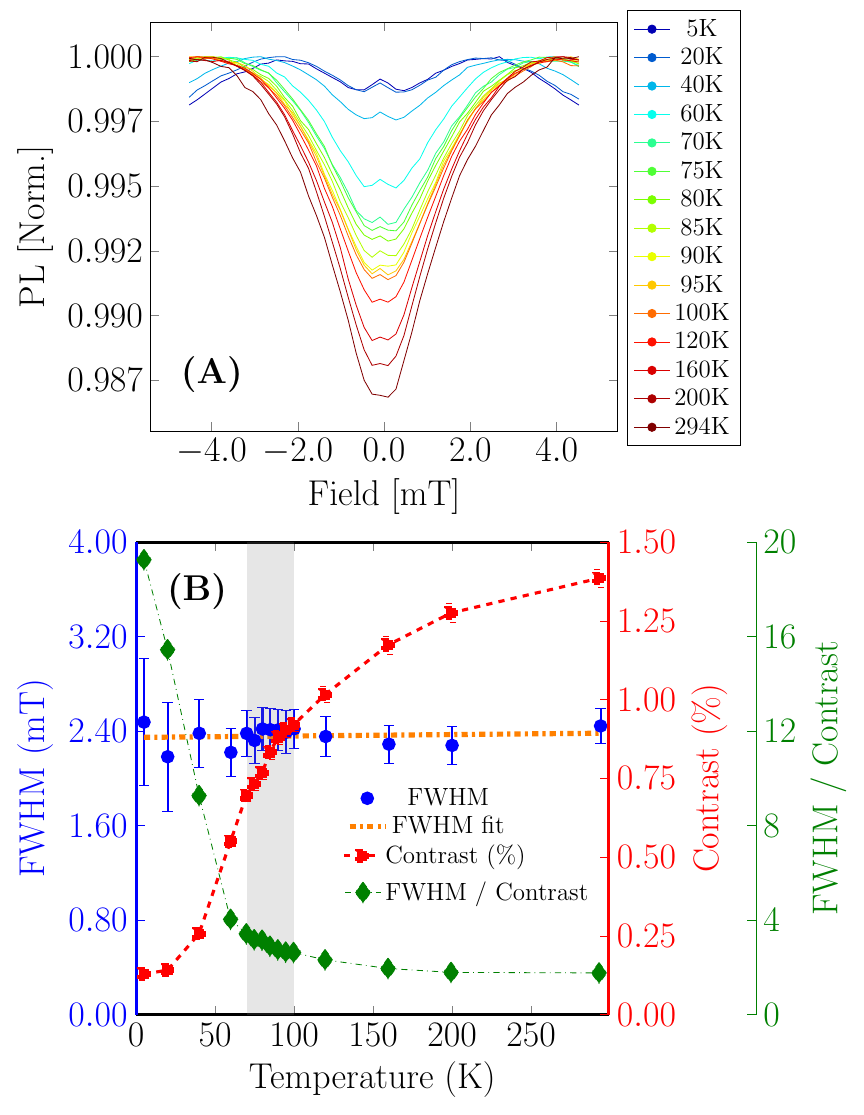}
    \caption{Cross-relaxation feature characterization for NDs. (A) The feature recorded at different temperatures; (B) contrast, linewidth, and the linewidth-to-contrast ratio of the cross-relaxation feature as a function of temperature. The gray area indicates the temperature region of interest for studying the superconductor. The temperature dependence of FWHM is fitted with linear dependence, The error bars are the fitting errors, which are not purely statistical.}
    \label{fig:linewidth_contrast}
\end{figure}

\section{Results and discussions}
\subsection{Temperature dependence of the cross-relaxation feature in NDs}
In this section, we outline the methodology used to characterize the temperature dependence of the cross-relaxation feature in NDs across relevant temperature ranges. 
We first characterized the temperature dependence of the cross-relaxation feature of NV centers in NDs deposited on a cover glass. A background magnetic field, applied perpendicular to the plane of the ND layer, was scanned from -6.5\,\si{\milli\tesla} to +6.5\,\si{\milli\tesla} in increments of 0.2\,\si{\milli\tesla}, and the fluorescence profile was recorded for various temperatures. Figure \ref{fig:linewidth_contrast}(a) shows the cross-relaxation feature recorded at various temperatures. Given that the contrast of the cross-relaxation feature is temperature dependent \cite{dhungel_near-zero-field_2024}, this step is crucial for determining the B-field modulation depth when probing the superconductor in the temperature range of interest.

 The full width at half maximum\,(FWHM) and contrast were extracted from 5\,K to 294\,K, with more measurements taken in the 70--100\,K temperature range with 5\,K steps, as illustrated in Fig.\,\ref{fig:linewidth_contrast}(b). The extracted dependences were fitted using a Lorentzian function to determine these parameters. The linewidth is relatively constant across all temperatures, and the contrast increases with increasing temperature.  This behavior is similar to that previously reported for bulk diamond \cite{dhungel_near-zero-field_2024}. The variation in contrast is less than 0.2\% between 70\,K to 100\,K, and the ratio of linewidth and contrast, and therefore, the sensitivity of the technique is about a factor of two worse within this temperature range compared to room temperature.

\subsection{Estimation of critical transition temperature of YBCO}\label{characterization}
The contrast of the cross-relaxation feature is above 1\% above 70\,K and the fluorescence change at 1\,mT is around 0.5\%. This change in fluorescence level is sufficient to observe the Meissner effect on the NDs when the magnetic field is modulated. 

\begin{figure}
\centering
    \includegraphics[width=0.47\textwidth]{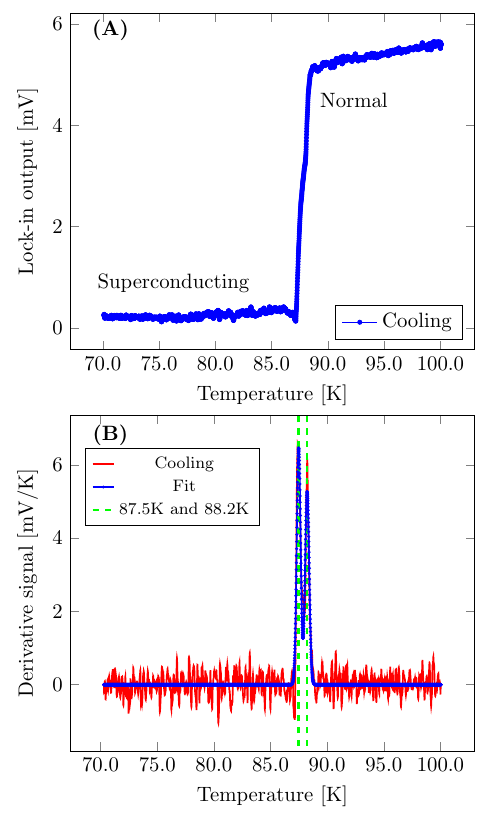}
    \caption{Lock-in amplifier output of the amplitude modulated fluorescence signal as a function of temperature. The magnetic field is modulated between 0 and 1\,mT with 3\,Hz and the resulting $R$ output from the lock-in amplifier is continuously acquired. The amplitude level is nearly zero below the superconducting transition temperature, and the level rises when the temperature reaches the transition temperature. 
    Figure (B) shows the derivative of the acquired signal, which is fitted with a double Gaussian function. The peak position obtained from the fit is taken as a transition temperature.}
    \label{fig:Transition_temp}
\end{figure}

To measure the critical temperature using the cross-relaxation feature, a 3\,Hz square-wave modulated 1\,mT magnetic field was applied. A control signal from the function generator serves as a reference for the lock-in amplifier. The modulated fluorescence response signal from the APD was fed to the input of the lock-in amplifier. The sample was initially cooled to 100\,K without a magnetic field and then further cooled to 70\,K at a rate of 1\,K per minute in the presence of a modulated magnetic field.

The output voltage amplitude $R$ at the output of the lock-in amplifier is relatively high ($\approx$\,6\,mV; see Fig.\,\ref{fig:Transition_temp}(A)) above the superconducting transition temperature as the NDs are exposed to the modulated magnetic field, which penetrates the normal (non-superconducting) state of the material. Upon cooling through the transition temperature, the measured signal drops sharply and remains close to zero below 88\,K, a consequence of the Meissner effect. In this regime, the superconductor completely expels the applied modulated field, see Fig.\,\ref{fig:setup}(B), resulting in a constant fluorescence level from the NDs. 

The derivative of the recorded data was smoothed using a moving average over 20 data points to suppress noise while preserving the dominant features. The resulting curve was then fitted with a double Gaussian function, and the center positions of the fitted peaks were taken as the effective transition temperatures. The gradual drop of signal above $T_c$ was not considered for the fitting procedure, as it does not influence the fit accuracy. Two distinct transition peaks were identified at 87.5\,K (FWHM = 0.4\,K) and 88.2\,K (FWHM = 0.4\,K), as shown in Fig.\,\ref{fig:Transition_temp}\,(B). This splitting is most likely due to oxygen inhomogeneity within the sample\,\cite{khadzhai2019annealing}. However, both values fall within the transition temperature range (87.5\,K) specified by the manufacturer.

\subsection{Meissner effect and magnetic field penetration}

\begin{figure*}[t]
    \centering
    \includegraphics[width=0.9\textwidth]{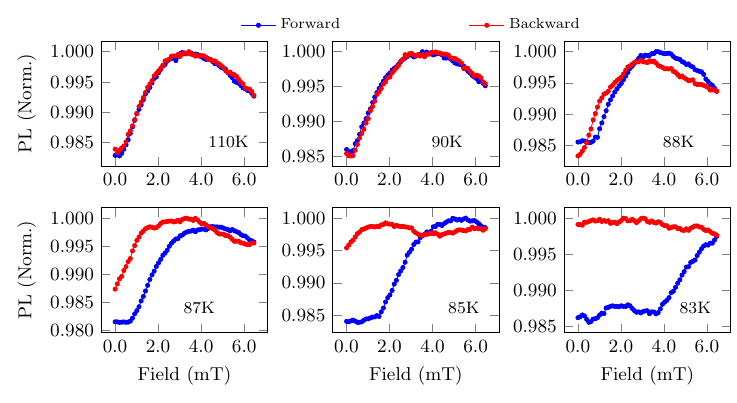}
    \caption{Plots for the local transition from the Meissner state to intermediate state at the center of the superconductor. The magnetic field is scanned in both forward (blue) and backward (red) direction at various temperatures including above and below the transition temperature.}
    \label{fig:field_scan_center_edge}
\end{figure*}

\begin{figure*}[t]
    \centering
    \includegraphics[width=0.9\textwidth]{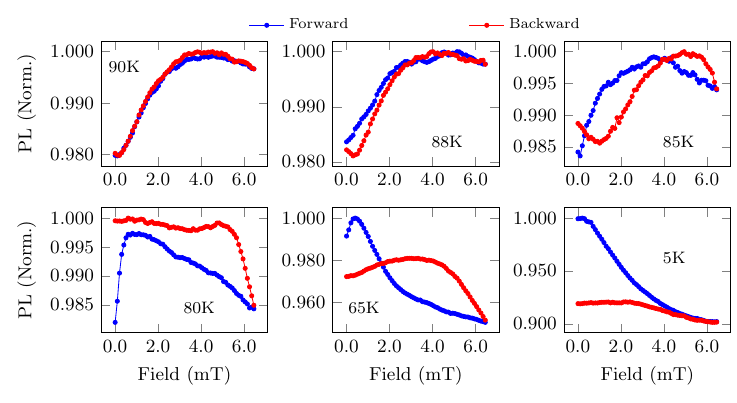}
    \caption{The amplification of the magnetic field at the superconductor edge. The external field was scanned in both forward (blue) and backward (red) directions across a range of temperatures, spanning regimes both above and below the superconducting transition temperature.}
    \label{fig:field_scan_edge}
\end{figure*}

In the Meissner-London state (no vortices), the superconductor expels the applied perpendicular magnetic field $H_0$, with screening currents flowing near the surfaces to achieve $B \approx 0$ inside. In this state, the London equation (see, for example, \cite{Annett2004}) provides an accurate description of the magnetic field expulsion inside the superconducting region. Exact analytical solutions of the London equation are known for specific geometries: infinite rectangular slab, a cylinder in perpendicular field, a sphere\,\cite{Prozorov2000, FIOLHAIS2014}. For the given geometry, a square platelet in our case, with side length $L = 5$ mm and thickness $d = 200$ nm, the aspect ratio $L/d = 25000$ is large, allowing for a 2D thin-film approximation to be used. Since the screening currents for a square geometry come from four sides, the actual suppression of $B$ is typically stronger (effectively $B = 0$) at the center. For $\lambda \ll \it{a}$ (thin film strip of width 2$\it{a}$), the field penetrates only near the edges and corners, focusing the flux at the edge. In such a scenario, the NV centers near the edge of the superconductor sense an enhanced magnetic field relative to NV centers at the center, with an edge-to-center enhancement factor $\eta = B_{edge}/B_{center}$. To observe the field enhancement, we monitored the fluorescence change from the NV centers for temperatures above and below the critical temperature at two different locations: at the sample center\,(Fig.\,\ref{fig:field_scan_center_edge}) and near the edge (Fig.\,\ref{fig:field_scan_edge}) of the superconductor. The superconductor was first cooled in a zero magnetic field and stabilized at a set temperature before the applied field was swept. For a fixed location, at a given temperature, the background magnetic field, applied perpendicular to the plane of the ND layer, thus along the c-axis of the SC sample, was swept linearly from 0\,\si{\milli\tesla} to 6.5\,\si{\milli\tesla} (forward) and 6.5\,\si{\milli\tesla} to 0\,\si{\milli\tesla} (backward). At each field value, the integrated fluorescence signal was recorded using a photodiode. This fluorescence serves as an optical probe: below the lower critical field \( H_{c1}\) the Meissner effect expels the applied field from the superconductor, while above \( H_{c1}\), vortex entry into the sample produces a measurable change in the fluorescence level. The penetration field was identified from the point at which the cross-relaxation feature in the fluorescence response changed shape, corresponding to the first penetration of the magnetic field into the sample. This microwave-free optical method provides a direct means to extract the penetration field from which the vortices start forming and offers potential for studying superconducting properties.

\textbf{Center of the SC} (Fig.\,\ref{fig:field_scan_center_edge}):
The strength of the lower critical field, where the onset of flux penetration occurs, varies significantly with temperature and is directly related to the transition temperature of the superconductor. At high temperatures above \( T_c \) (e.g., 110\,K and 90\,K), the fluorescence signal follows the expected structure [Fig.\,\ref{fig:linewidth_contrast}\,(a)], and does not exhibit changes between forward and backward scans of the applied magnetic field, consistent with the absence of superconductivity. At 88\,K, near the critical temperature, we observe superconductivity (flux expulsion) up to approximately 0.4\,mT, which is followed by a change in the fluorescence level at higher fields, marking the onset of flux penetration, indicating the superconductor transitioning from a pure Meissner state to a mixed vortex state. This is because the scanning field exceeds the lower critical field. Further below \( T_c \), the penetration threshold increases: at 83\,K we find a change in the fluorescence level at 3.64$\pm$0.04\,mT, marked by a >1\% increase in fluorescence. The fluorescence level for the backward field sweep is constant, as the magnetic flux is trapped inside the superconductor.
 
The hysteresis between forward and backward sweeps at lower temperatures is explained by vortex pinning: once vortices form in fields that exceed the penetration field, they become anchored to structural defects in the crystal lattice \cite{chen2024revealing, mahato2024pinning, ishida2019unique}. Because the pinning force is largely independent of the external field, vortices remain trapped even when the field is reduced, producing a persistent signal. Below \( T_c\), our measurements thus capture the magnetic contribution of these immobile vortices. It is important to note that the width and contrast of the feature depend on the density of NVs in ND, the intensity of the laser and the temperature\,\cite{dhungel_near-zero-field_2024}.  Optimal contrast is achieved in regions of the YBCO film where ND agglomeration is minimized and spatial homogeneity is maximized, thereby reducing the standoff distance between the nanodiamonds and the superconducting layer.

 \textbf{Edge of the SC} (Fig.\,\ref{fig:field_scan_edge}):
 The fluorescence plots in Fig.\,\ref{fig:field_scan_edge} shows the response from NDs at the edge of the YBCO film, provides a strong evidence for both field enhancement and the vortex pinning.

 Compared to the fluorescence response at the center of the superconductor described above (Fig.\,\ref{fig:field_scan_center_edge}), below $T_c$ we observe reduced fluorescence in the backward field scans, consistent with an amplification of the field measured by the NDs. 
This observed behavior is an interplay between field expulsion (Meissner state), vortex penetration along the edges, and flux pinning. At temperatures above the critical temperature (90\,K), the YBCO film is in normal state similar to the sample center. As the magnetic field is swept, the local field activity at the edge due to the flux focusing causes the applied magnetic field to reach the lower critical field at a much lower value than the field required at the center.
This effect could be understood by a direct comparison of the data from Fig.\,4 that shows the measurement at the center of the sample. For example, at 83\,K, the center shows flux penetration at 3.64\,mT, and in contrast the edge shows a clear drop in the fluorescence at a much lower field, approximately 0.5\,mT.

An exact quantitative determination of the local field amplification below \( T_c \) requires proper calibration of the detection scheme, including reference measurements on bare NDs without the superconductor for direct comparison, which would require higher magnetic field amplitudes than our setup can provide. However, the shape of the observed amplification features can still be compared with reports in the literature \cite{rodzon2025temperature,wickenbrock2016microwave}. Consistently, our measurements reveal a noticeable enhancement of the apparent field as the temperature decreases (Fig.\ref{fig:field_scan_edge}).
This behavior can be attributed to flux-focusing and demagnetization effects that arise near the superconductor edge, which become more pronounced as superconducting screening currents strengthen at lower temperatures.

\textbf{Present limitations:}
In this analysis, geometric effects are neglected and the penetration field is assumed to be equal to \(H_{c1}\) for sample dimensions of 5\,mm$\times$5\,mm$\times$200\,nm, measured at the sample center. For a precise estimation of \(H_{c1}(0)\) and, consequently, the superconducting coherence length, the penetration field should ideally be measured at temperatures far below the critical temperature. Such measurements require the application of higher magnetic fields ($\approx$ 40\,mT) whereas the maximum field attainable in the present setup is limited to 7\,mT.

The current study uses a widefield measurement approach with a region of interest measuring 120\,μm × 120\,μm. As a result, the penetration field reported and the field enhancement observed represent an average over a relatively large area, incorporating the effects from spatial variations of the screening currents, geometric imperfections, inhomogeneity of ND distribution, and laser power. These inhomogeneities, intrinsic to the larger sampled regions, could be minimized in future work by employing confocal microscopy to probe smaller, well-defined local regions \cite{joshi_measuring_2019}.


\section{Conclusions}
This study demonstrates a microwave-free method that utilizes nanodiamond to quantitatively investigate superconducting properties. Using cross-relaxation features, we successfully measured $T_c$ of YBCO and the penetration field at different temperatures near $T_c$ with minimal invasiveness and without microwaves. The insensitivity of the technique to surface roughness and the avoidance of microwave interference make it easier to implement than existing NV-based approaches. To improve, a confocal microscope can be used instead of our widefield setup for localized measurement of $T_c$ and \(H_{c1}\). However, the use of the widefield technique opens up the possibility of imaging superconducting properties over the field of view. Future applications could extend to vortex dynamics, thin-film heterostructures, and topological superconductors, offering a convenient probe of various phenomena in quantum materials. This work underscores the potential of NV magnetometry to bridge macroscopic measurements and microscopic physics in superconductivity research.

\section*{Acknowledgements} 
The authors acknowledge helpful discussions with Dieter  Koelle and Patrick Maletinsky. This work was supported by the European Commission’s Horizon Europe Framework Program under the Research and Innovation Action MUQUABIS GA no. 101070546, by the German Federal Ministry of Education and Research (BMBF) within the Quantumtechnologien program (DIAQNOS,
project no. 13N16455) and by the Deutsche Forschungsgemeinschaft (DFG, German Research Foundation) in the framework to the collaborative research center "Defects and Defect Engineering in Soft Matter" (SFB1552) under Project No. 465145163. This research was funded in part by the National Science Centre, Poland, Grant No. 2020/39/I/ST3/02322, 2024/53/N/ST3/04343 and Grant Nos. 2021/03/Y/ST3/00185, 2023/05/Y/ST3/00135 within the QuantERA II Programme that has received funding from the European Union’s Horizon 2020 research and innovation programme under Grant Agreement No. 101017733. For the purpose of Open Access, the author has applied a CC-BY public copyright licence to any Author Accepted Manuscript (AAM) version arising from this submission. The UJ authors acknowledge the use of Tidy3D (Flexcompute Inc.) for modeling the NDs optical emission.

\bibliography{new.bib}
\end{document}